# Comment on "An arbitrated quantum signature scheme with fast signing and verifying"

Yi-Ping Luo[1] and Tzonelih Hwang[*]

*Department of Computer Science and Information Engineering, National Cheng Kung University, No. 1, University Rd., Tainan City, 70101, Taiwan, R.O.C.*

[1] yiping@ismail.csie.ncku.edu.tw

[*] hwangtl@ismail.csie.ncku.edu.tw

## Abstract

Recently, Liu et al. (Quantum Inf Process (2014) 13:491–502) proposed an arbitrated quantum signature (AQS) scheme, where a signature receiver (Bob) can verify the signer's signature through the help of a trusted arbitrator. However, this paper shows that a malicious Bob can perform the existential forgery of the signature under the chosen message attack without being detected.

## 1  Introduction

Arbitrated quantum signature (AQS) is one of the imperative research topics in quantum cryptography which guarantees the authentication of identities and the integrity of the classical messages or quantum states over insecure quantum channels [1-3]. Usually, in an AQS scheme, a trusted arbitrator helps a receiver to validate the legitimacy of the signature. Similar to the classical digital signature, an AQS should satisfy the following security requirements [3]:

1. **Unforgeability:** Neither the signature receiver nor an attacker can forge a signature or change the content of a signature.
2. **Non-repudiation:** After signing a valid signature, a signer should not be able to deny that.

In 2001, Gottesman and Chuang [4] firstly brought out the idea of designing an AQS scheme based on fundamental principles of quantum physics. After that, various AQS schemes have been proposed [1-3, 5-19]. Recently, Liu et al. proposed an AQS scheme with fast signing and verifying technique [20], where a new quantum one-time pad (QOTP) called D-QOTP (QOTP using decoy states) is designed to avoid being forged and disavowed [18, 19]. However, in this paper, we show that a malicious receiver, Bob, can perform the existential forgery of the signature under the chosen message attack by using several valid quantum message and signature pairs without being detected. Therefore, the requirements of unforgeability and non-repudiation are not satisfied in Liu et al.'s AQS scheme.

The rest of this paper is organized as follows. Section 2 reviews Liu et al.'s AQS scheme. Section 3 describes the existential forgery in Liu et al.'s scheme. Section 4 summarizes the result.

## 2  Review of Liu et al.'s AQS Scheme

In this section, at first we describe the technique of the D-QOTP (i.e., quantum one-time pad using decoy states) algorithm which used in Liu et al.'s AQS scheme. Subsequently, we briefly review of Liu et al.'s AQS scheme.

## 2.1 The D-QOTP Algorithm

Suppose $K$ denotes the secret key shared between the sender and the receiver. The quantum message $|P\rangle$ can be encrypted to $|C\rangle$ as follows, where $|P\rangle = \otimes_{i=1}^{n}|p_i\rangle$, $|p_i\rangle = \alpha_i|0\rangle + \beta_i|1\rangle$, $\alpha_i, \beta_i \in \mathbf{C}$, $|\alpha_i|^2 + |\beta_i|^2 = 1$, and $1 \leq i \leq n$.

**Encryption Algorithm of D-QOTP**

E1. Split $K$ into $2^{t+1}-2$ substrings $Q = K_{2(L1)}; K_{2^2(L1)}, K_{2^2(L2)};\ldots; K_{2^t(L1)},\ldots, K_{2^t(L2^{t-1})}. K_{2(R1)}; K_{2^2(R1)}, K_{2^2(R2)};\ldots; K_{2^t(R1)},\ldots, K_{2^t(R2^{t-1})}$ (more details please see [20]), where $2^{t+1} \geq n+3$.

E2. Every substrings $K_{2^i(L2^j)}$ and $K_{2^i(R2^j)}$ can be interpreted as decimal integers $i_{Lj}$ and $i_{Rj}$, respectively, where $1 \leq i \leq t$ and $1 \leq j \leq 2^{i-1}$. That is, $(Q)_{10} = 1_{L1}; 2_{L1}, 2_{L2};\ldots; t_{L1},\ldots, t_{L2^{t-1}}. 1_{R1}; 2_{R1}, 2_{R2};\ldots; t_{R1},\ldots, t_{R2^{t-1}}$.

E3. $R$ is the quantum sequence in the states of the loop of $\{|0\rangle, |1\rangle, |+\rangle, |-\rangle\}$, i.e., $R = (|0\rangle, |1\rangle, |+\rangle, |-\rangle; |0\rangle, |1\rangle, |+\rangle, |-\rangle; \ldots)$.

E4. $|C\rangle = E_K(|P\rangle)$, where the decoy state of $|R\rangle$ is inserted into $|P\rangle$ to form $|C\rangle$ based on $(Q)_{10}$.

Finally, we can get the output $|C\rangle = E_K(|P\rangle)$ from E1 to E4.

Based on the cipher-text $|C\rangle$ and the secret key $K$, the decryption algorithm is described as follows.

**Decryption Algorithm of D-QOTP**

D1. The same as Step E1, split $K$ into the string $Q$.

D2. The same as Step E2 to obtain the decimal integer string $(Q)_{10}$.

D3. The same as Step E2 to construct $|R\rangle$.

D4. Extract the decoy states from $|C\rangle$ based on $(Q)_{10}$, which is denoted as $|R\rangle'$. Subsequently, measure $|R\rangle'$ with the bases which are indicated in $|R\rangle$. Verify the measurement result to check the eavesdropping and the integrity of $|P\rangle$. If there exists an eavesdropping, this session will be aborted and restarts the protocol again.

Finally, we can decrypt $|C\rangle$ to obtain $|P\rangle = D_K(|C\rangle)$.

## 2.2 A Brief Review of Liu et al.'s AQS scheme

Here, a signer Alice wants to sign on the quantum message $|P\rangle$ and transmits it to the signature receiver, Bob. Subsequently, Bob can verify Alice's signature with the help of a trusted arbitrator, Trent. Liu et al.'s AQS scheme is composed of three phases: the initializing phase, the signing phase, and the verifying phase.

**Initializing phase**

**Step I1.** Trent shares the secret keys $K_A$ and $K_B$ with Alice and Bob, respectively, through the unconditionally secure quantum key distribution protocols, where $K_A \in \{0,1\}^{L_A}$, $K_B \in \{0,1\}^{L_B}$, $L_A \geq \left[\frac{n}{2}\right] + 2$, and $L_B \geq \left[\frac{n + L_A}{2}\right] + 2$.

**Step I2.** Alice, Bob, and Trent choose a loop sequence $|R\rangle$ from $\{|0\rangle, |1\rangle, |+\rangle, |-\rangle\}$ as a set of the decoy states.

**Signing phase**

**Step S1.** Alice prepares the quantum message $|P\rangle$, in which if $|P\rangle$ is composed of known quantum states, then arbitrary copies of $|P\rangle$ can be produced. If $|P\rangle$ is composed of unknown quantum states, then there need at least three copies of $|P\rangle$, i.e. $|P\rangle_1$, $|P\rangle_2$, and $|P\rangle_3$, where $|P\rangle_1 = |P\rangle_2 = |P\rangle_3$.

**Step S2.** Follow Step E1, Alice obtains $Q_A$.

**Step S3.** From Step E2, Alice obtains $(Q_A)_{10}$.

**Step S4.** Alice generates her quantum signature $|S\rangle = E_{K_A}(|P\rangle_3)$ based on $|R\rangle$ and $(Q_A)_{10}$.

**Step S5.** Alice sends $|S\rangle \otimes |P\rangle_1 \otimes |P\rangle_2$ to Bob.

**Verifying phase**

**Step V1.** Upon receiving the quantum sequence, Bob compares whether $|P\rangle_1 = |P\rangle_2$ by using quantum fingerprinting [21]. If the comparison result is negative, then Bob aborts this transmission and informs Alice to restart the scheme. Otherwise, Bob splits $K_B$ into $Q_B$ and then obtains $(Q_B)_{10}$ by using the same way in Step E1 and E2.

**Step V2.** Bob transforms $|P\rangle_2$ into $|T\rangle = E_{K_B}(|P\rangle_2)$ by using $|R\rangle$ and $(Q_B)_{10}$.

**Step V3.** Bob keeps $|P\rangle_1$ and $(Q_B)_{10}$, and then sends $|S\rangle \otimes |T\rangle$ to Trent.

**Step V4.** Trent splits $K_A$ and $K_B$ into $Q_A$ and $Q_B$ to obtain $(Q_A)_{10}$ and $(Q_B)_{10}$, respectively, which is identical with Step D1~D3. Subsequently, Trent extracts the decoy states from $|S\rangle$ and $|T\rangle$ based on $(Q_A)_{10}$ and $(Q_B)_{10}$ to obtain $|P'\rangle_3$ and $|P'\rangle_2$, respectively.

**Step V5.** Following Step D4, Trent measures the extracted decoy states $|S\rangle \backslash |P'\rangle_3$ and $|T\rangle \backslash |P'\rangle_2$ with the bases which are indicated in $((Q_A)_{10}, |R\rangle)$ and $((Q_B)_{10}, |R\rangle)$, respectively. In this case, Trent can check whether there exists an eavesdropper or not. Besides, Trent can also check existence of any forgery attack from the measurement results. Subsequently, Trent compares $|P'\rangle_3$ and $|P'\rangle_2$. If $|P'\rangle_3 \neq |P'\rangle_2$, Trent aborts this communication and the scheme needs to be restarted. Otherwise, Trent continues to the next step.

**Step V6.** Trent transforms $|P'\rangle_2$ to $|T\rangle = E_{K_B}(|P'\rangle_2)$ based on $|R\rangle$ and $(Q_B)_{10}$. After that, Trent transforms $|P'\rangle_3$ to $|S\rangle_T = E_{K_B}(E_{K_A}(|P'\rangle_3))$ based on $((Q_A)_{10}, |R\rangle)$ and $((Q_B)_{10}, |R\rangle)$. Trent sends $|T\rangle \otimes |S\rangle_T$ to Bob.

**Step V7.** Bob extracts and measures the decoy states from $|T\rangle$ and $|S\rangle_T$ based on $((Q_B)_{10}, |R\rangle)$, respectively, and denotes the rest particles of $|T\rangle$ and $|S\rangle_T$ as $|P''\rangle_2$ and $|S''\rangle$. Bob verifies the measurement results, if there exists an eavesdropping, then Bob rejects the signature. Otherwise, Bob goes to the next step.

**Step V8.** Bob compares $|P''\rangle_2$ with his retained $|P\rangle_1$. If $|P''\rangle_2 = |P\rangle_1$, then Bob accepts $|S''\rangle$. Otherwise, Bob rejects $|S''\rangle$.

## 3  The Existential Forgery of Signature

In this section, we demonstrate that a malicious receiver, Bob, is able to perform the existential forgery of signer's signature without being detected as follows.

In their AQS scheme, Alice's quantum signature $|S\rangle$ is generated based on the loop sequence $|R\rangle$ and the secret key $K_A$. For singing the different quantum messages (i.e., $|P\rangle_A$ and $|P\rangle_B$, where $\||P\rangle_A\| = \||P\rangle_B\| = n$, and $|P\rangle_A \neq |P\rangle_B$), Alice generates different quantum signatures ($|S\rangle_A$ and $|S\rangle_B$) based on the loop sequence $|R\rangle$ and the secret key $K_A$. However, because of the usage of the same secret key, the positions of the decoy states would always be the same in two different quantum signatures. A malicious receiver, Bob, may collect several quantum signatures in order to comprehend the positions of the quantum messages and the decoy states. Once the positions of the quantum message are revealed to *n*-bit length, Bob can modify the pair of the quantum message and the quantum signature together by using unitary operations without being detected. Therefore, their scheme cannot satisfy the requirements of unforgeability.

For example, suppose there are two quantum messages $|P\rangle_A = |0\rangle|0\rangle|0\rangle|0\rangle$, $|P\rangle_B = |1\rangle|1\rangle|1\rangle|1\rangle$, $\||P\rangle_A\| = \||P\rangle_B\| = 4$, $K = 1011$, and $|R\rangle = (|0\rangle, |1\rangle, |+\rangle)$. We can divide $K$ and obtain $Q = \left(K_{2(L1)}; K_{2^2(L1)}, K_{2^2(L2)}, K_{2(R1)}; K_{2^2(R1)}, K_{2^2(R2)}\right) = (10;1,0.11;1,1) = (10;1.11;1)$, where the zero position is ignored and the exclusion principle is used in the last equation. Subsequently, $(Q)_{10} = (2;1.3;1)$. For the quantum messages $|P\rangle_A$ and $|P\rangle_B$, the resulting signatures are $|S\rangle_A = (|1\rangle, |p_1\rangle_A, |0\rangle, |p_2\rangle_A, |+\rangle, |p_3\rangle_A, |p_4\rangle_A, |0\rangle) = (|1\rangle, |0\rangle, |0\rangle, |0\rangle, |+\rangle, |0\rangle, |0\rangle, |0\rangle)$ and $|S\rangle_B = (|1\rangle, |p_1\rangle_B, |0\rangle, |p_2\rangle_B, |+\rangle, |p_3\rangle_B, |p_4\rangle_B, |0\rangle) = (|1\rangle, |1\rangle, |0\rangle, |1\rangle, |+\rangle, |1\rangle, |1\rangle, |0\rangle)$, respectively. Now, Bob compares $|S\rangle_A^i$ and $|S\rangle_B^i$ by using quantum

fingerprinting, where $1 \leq i \leq 8$. Bob can obtain the comparison results of $|S\rangle_A^2 \neq |S\rangle_B^2$, $|S\rangle_A^4 \neq |S\rangle_B^4$, $|S\rangle_A^6 \neq |S\rangle_B^6$, $|S\rangle_A^7 \neq |S\rangle_B^7$. Since the number of $|S\rangle_A^i \neq |S\rangle_B^i$ is equal to the length of the quantum message (i.e., 4), therefore, Bob can perform any unitary operation on the new quantum message and the corresponding quantum signature in the positions $(2,4,6,7)$ without being detected. Due to this attack, the signer, Alice, can later deny that she has signed a new quantum message. Therefore, Liu et al.'s AQS scheme cannot satisfy the requirements of the unforgeability as well as non-repudiation.

## 4 Conclusions

In this article, we have pointed out that Liu et al.'s AQS scheme suffers from the existential forgery of the signature under the chosen message attack performed by a signature receiver, Bob. The possible way to resolve this issue is that, the signer (Alice) has to share a new secret key with the arbitrator (Trent), which requires a QKD protocol to perform between them. However, this approach is not feasible for a signature scheme and that also impair the efficiency of the protocol. Therefore, how to design an AQS scheme with the feature of key re-usability would be an interesting research topic.

## Acknowledgment

We would like to thank the Ministry of Science and Technology of Republic of China for financial support of this research under Contract No. MOST 104-2221-E-006-102 -.